\documentclass[onecolumn]{revtex4}

\usepackage{graphicx}
\usepackage{dcolumn}
\usepackage{bm}

\input epsf

\begin{document}

\title{\Large THERMODYNAMICS OF THE UNIVERSE FILLED WITH PERFECT FLUID HAVING VARIABLE EQUATION OF STATE}

\author{\bf~Nairwita~Mazumder\footnote{nairwita15@gmail.com}, Subenoy~Chakraborty\footnote{schakraborty@math.jdvu.ac.in}.}

\affiliation{Department of Mathematics,~Jadavpur
University,~Kolkata-32, India.}

\date{\today}

\begin{abstract}
We study the thermodynamics of the universe containing perfect
fluid with equation of state $p=\omega\rho,\omega$ is variable.
Here we choose $\omega$ to be a function of the red shift
variable $z$ and four different choices of $\omega$ has been
investigated. Also the laws of thermodynamics are examined
considering the universe bounded by the horizon (apparent) as a
thermodynamical system.
\end{abstract}

\maketitle

\section{\normalsize\bf{Introduction}}
In the semi-classical description of black hole physics it is
found that a black hole behaves as a black  body and emits thermal
radiation. The temperature (known as Hawking temperature) and the
entropy are proportional to the surface gravity at the horizon and
area of the horizon [1,2] respectively. The Hawking~
temperature,entropy and mass of the black hole satisfy the first
law of thermodynamics [3]. As the temperature and entropy are
determined by purely geometric quantities(namely surface gravity
and horizon area respectively),i.e. characterized by the
space-time geometry and hence by Einstein field equations,so it is
natural to speculate some relationships between black hole
thermodynamics and Einstein equations. Jacobson [4] showed that
Einstein equations can be derived from the first law of
thermodynamics: $\delta$Q=Tds for all local Rindler causal
horizons with $\delta$Q and T as the energy flux and unruh
temperature seen by an accelerated observer just inside the
horizon. For a general static spherically symmetric space-time,
Padmanabhan [5] was able to derive the first law of
thermodynamics on the horizon, starting
from Einstein equations.\\

Subsequently, this equivalence between Einstein equation and
thermodynamical laws has been generalized in the context of
cosmology. If we assume the universe as a thermodynamical system
and consider at the apparent horizon $R_{A}$ the Hawking
temperature ~$T_{A}=\frac{1}{2\pi R_{A}}$ and the entropy
~$S_{A}=\frac{\pi{R_{A}}^{2}}G$~, then it was shown that the first
law of thermodynamics on the apparent horizon and the Friedmann
equations are equivalent [6]-one can be derived from other. Then
in higher dimensional space-time, the relation was established for
gravity with Gauss-Bonnet term and for the Lovelock gravity
theory[6,7]. As a result, it is speculated that such a deep
relationship between the thermodynamics at the apparent horizon
and the Einstein equations may give some clue on the properties of
dark energy.\\

In the present work, we study the thermodynamics of the universe
with matter in the form of perfect fluid having variable equation
of state in form:~$p=\omega\rho$. We consider $\omega$ as a
function of the red shift $z$ in the following forms of two index
parametrization[8]:{\it
(i)}~$\omega=\omega_{0}+\omega_{1}z$~~(linear redshift
parametrization),{\it
(ii)}~$\omega=\omega_{0}+\frac{\omega_{1}z}{(1+z)}$
(Chevallier-Polarski linear parametrization), {\it
(iii)}~$\omega=\omega_{0}+\frac{\omega_{1}z}{(1+z)^{2}}$
(Jassal-Bagla-Padmanavan parameterization),{\it
(iv)}~$\omega=\omega_{0}+\omega_{1}z$~~if
~$z<1$~,$\omega_{0}+\omega_{1}$~~if~$z\geq1$
(Upadhya-Ishak-Steinhardt parametrization). The thermal quantities
are expressed either as a function of volume or temperature and
due to adiabatic nature of the thermodynamical  system the entropy
turns out to be a constant. Subsequently , we examine the validity
of  the thermodynamical laws for  universe bounded by the
horizon(event or apparent).\\

\section{\normalsize\bf{General Thermodynamical Description}}

Let us consider a thermodynamical system bounded in a volume $V$
and suppose  $\rho$ ,$p$ and $ T $ are the energy density,
thermodynamical pressure and temperature of the fluid bounded by
the volume. Then from the first law of thermodynamics [9]

\begin{equation}
TdS=d(\rho V)+pdV=d[(\rho + p)V]-Vdp
\end{equation}

where ~$S$~ is he entropy of the thermodynamical system .Now the
integrability condition [9]

\begin{equation}
\frac{\partial^{2}S}{\partial T \partial V} =
\frac{\partial^{2}S}{\partial V \partial T}~,
\end{equation}

demands the following differential relation[10]

\begin{equation}
\frac{dp}{\rho + p}=\frac{dT}T ~.
\end{equation}

Combining (1) and (3) and integrating once we obtain (except for
an additive constant)

\begin{equation}
S=\frac{(\rho + p)V}T
\end{equation}

However, for adiabatic process entropy is constant and
consequently, the first law of thermodynamics (1) becomes the
conservation law

\begin{equation}
d[(\rho + p)V]=Vdp
\end{equation}

Note that one can obtain the relation (4) easily using the
conservation relation (5)into the integrability condition(3).
Hence for adiabatic process equation (4) may be considered as the
temperature defining equation. If we suppose our universe to be
homogeneous and isotropic FRW space-time with line element

\begin{equation}
ds^{2}=-dt^{2}+a^{2}(t)\left[\frac{dr^{2}}{1-kr^{2}}+~{r}^{2}d{{\Omega}_{2}}^{2}\right]
\end{equation}

then the Einstein's equations (known as Friedmann equations) and
the energy conservation law are

\begin{equation}
H^{2}=\frac{8\pi G \rho}3 - \frac{k}{a^{2}}~,
\end{equation}

\begin{equation}
\dot{H}=4\pi G (\rho + p) + \frac{k}{a^{2}}
\end{equation}

and

\begin{equation}
\dot{ \rho }+3H(\rho + p) = 0~,
\end{equation}

where ~$H=\frac{\dot{a}}a$~ is the Hubble parameter, $k$ is the
curvature scalar having values ~$0,\pm1$~ for three dimensional
space to be flat or to have positive or negative spatial curvature
and ~$d{{\Omega}_{2}}^{2}$~ is the metric on unit 2-sphere.Now we
consider our universe filled with perfect fluid having equation of
state $p=\omega\rho$ where $\omega$ is not constant but a
function of the red shift $z$  i.e. $\omega=\omega(z)$ .We choose
the following forms for $\omega(z)$ namely{\it
(I)}~$\omega=\omega_{0}+\omega_{1}z$~~,{\it
(II)}~$\omega=\omega_{0}+\frac{\omega_{1}z}{(1+z)}$,~~ {\it
(III)}~$\omega=\omega_{0}+\frac{\omega_{1}z}{(1+z)^{2}}$,~~{\it
(IV)}~$\omega=\omega_{0}+\omega_{1}z$~~if
~$z<1$~,$\omega_{0}+\omega_{1}$~~if~$z\geq1$ ~~where
~$\omega_{0}$~and~$\omega_{1}$~~are constants.\\

\textbf{Case I :~$\omega(z)=\omega_{0}+\omega_{1}z$~~}
     From the energy conservation equation (8) integrating once we have

\begin{equation}
\rho(z)=\rho_{0}e^{3\omega_{1}z}{(1+z)}^{3(1+\omega_{0}-\omega_{1})}
\end{equation}

where $ \rho_{0} $ an integration constant.As the system is an
adiabatic process so entropy is constant ($S_{0}$, say) and hence
from equation (4)

\begin{equation}
T(z)=T_{0}e^{3\omega_{1}z}{(1+z)}^{3(\omega_{0}-\omega_{1})}(1+\omega_{0}+\omega_{1}z)
\end{equation}

with $ T_{0}=\frac{\rho_{0}}{S_{0}} $, a constant. The squared
speed of sound $ {v_{s}}^{2} $ is given by

\begin{equation}
{v_{s}}^{2}(z)=\frac{\partial p} {\partial
\rho}=\omega+\frac{\omega_{1}(1+z)}{3(1+\omega)}
\end{equation}

Finally, the heat capacity is calculated to be

\begin{equation}
C_{v}(z)=V\frac{\partial \rho} {\partial T} = \frac
{3S_{0}}{1+z}{\left[~{\frac{3 \omega}{1+z}+
\frac{\omega_{1}}{1+\omega}}~\right]}^{-1}
\end{equation}

\begin{figure}
\includegraphics[height=1.5in]{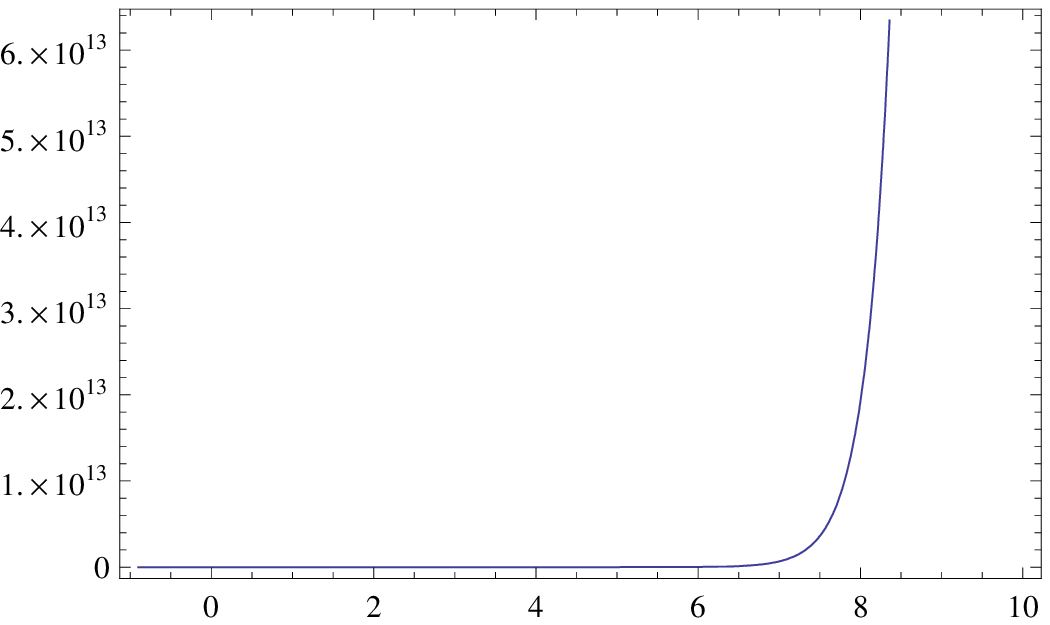}~~~
\includegraphics[height=1.5in]{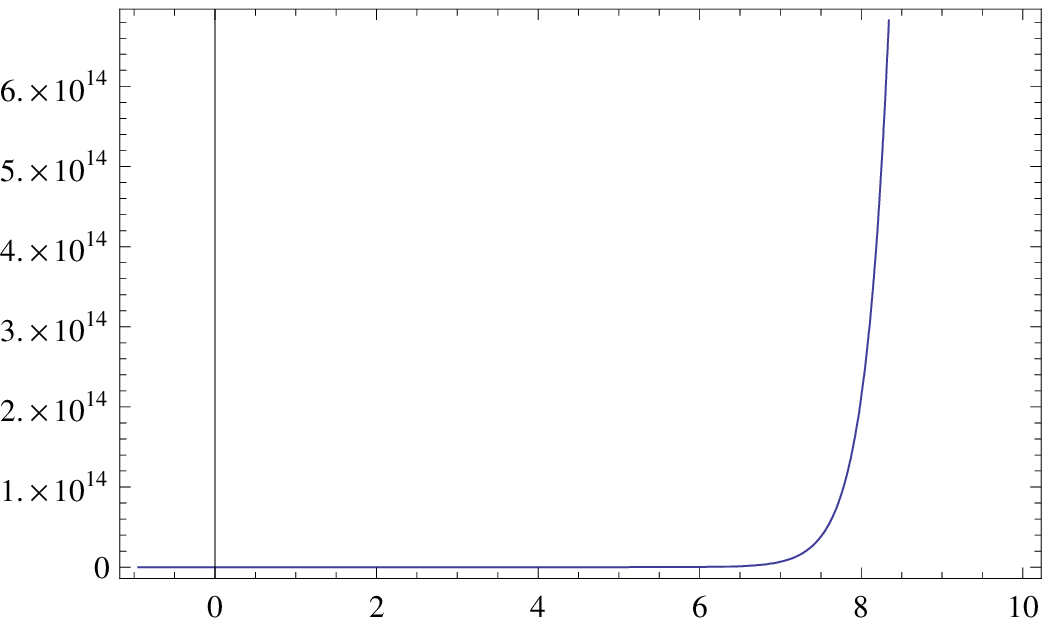}~~~
\\
\vspace{1mm}
Fig.1(a)~~~~~~~~~~~~~~~~~~~~~~~~~~~~~~~~~~~~~~~~~~~~~~~~Fig.1(b)\\
\includegraphics[height=1.5in]{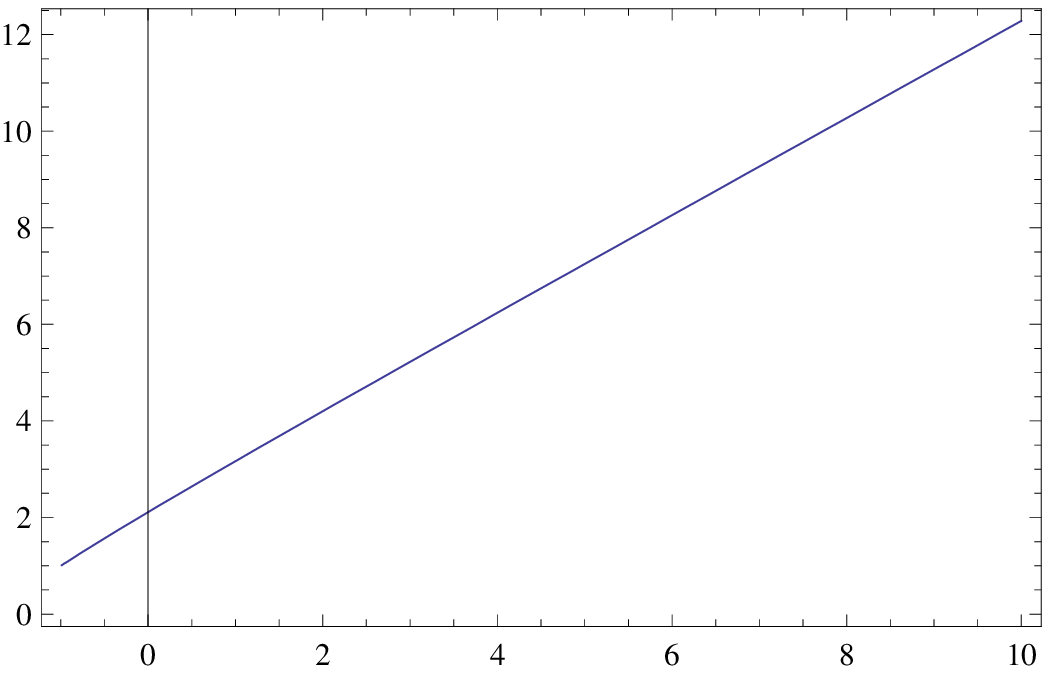}~~~
\includegraphics[height=1.5in]{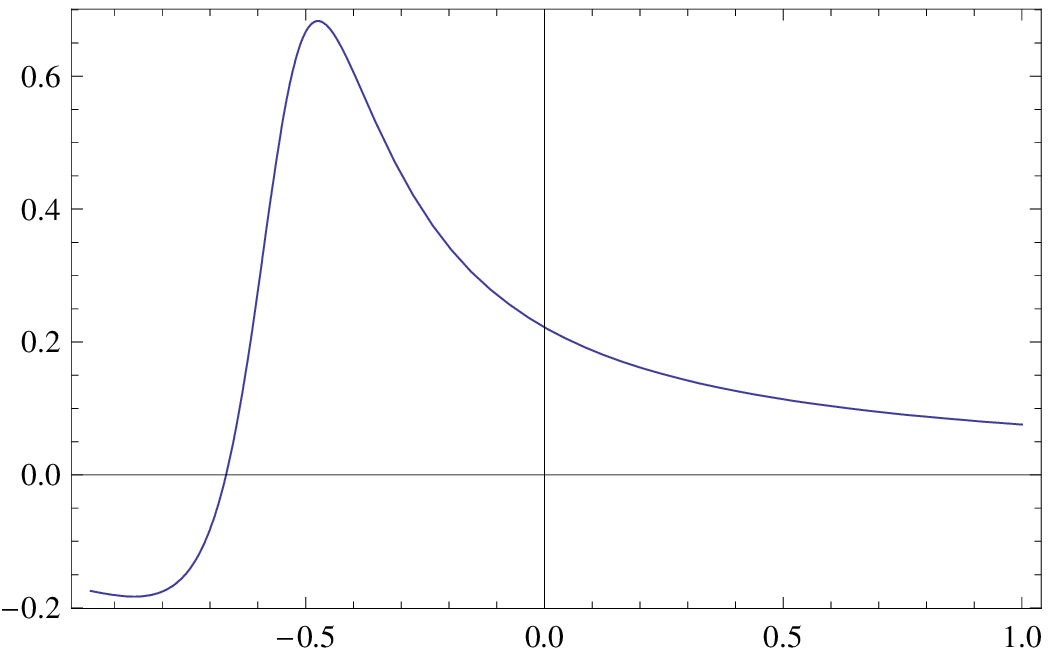}~~~
\\
\vspace{1mm}
Fig.1(c)~~~~~~~~~~~~~~~~~~~~~~~~~~~~~~~~~~~~~~~~~~~~~~~~Fig.1(d)\\

\vspace{5mm} Fig. 1(a),1(b),1(c) shows the variation of the
energy density,temperature and velocity of sound with the red
shift factor (given by eq.(10,11,12)) respectively where
 ~$\omega_{0}>\omega_{1} $~~ , here the
values of $\omega_{0}=2$ and $\omega_{1}=1$ . Fig.1(d) shows the
variation of heat capacity with red shift factor for
~$\omega_{0}<\omega_{1} $~~ with values of $\omega_{0}=1$ and
$\omega_{1}=3$. Here~$\rho_{0}=1,S_{0}=1$~\hspace{1cm}
\vspace{6mm}
\end{figure}

\textbf{Limiting Values : }\\

{\it(i)}~$ a \rightarrow 0$ i.e. $z \rightarrow \infty$~:~(from
fig.1)\\
 $\rho \rightarrow \infty $,$p\rightarrow \infty$,$T
\rightarrow
\infty $,$v_{s} \rightarrow \infty$ and $C_{v} \rightarrow 0$.\\

{\it(ii)}~ $ a \rightarrow \infty $ i.e.$z \rightarrow -1$~:~\\

~~{\it(a)}~~$\rho \rightarrow 0 $, $p\rightarrow 0 $~and ~$T
\rightarrow 0 $~ if~$\omega_{0}>\omega_{1} $~~\\

~~{\it(b)}~~$\rho \rightarrow \infty $ , $p\rightarrow {-\infty}
$~and~ $T \rightarrow \infty $~ if $\omega_{0}<\omega_{1}<1+\omega_{0} $~~\\

~~{\it(c)}~~$\rho \rightarrow \rho_{0}e^{-3\omega_{1}} $ and
$p\rightarrow {-\rho_{0}e^{-3\omega_{1}}} $ if
$1+\omega_{0}=\omega_{1} $~~ and~~$T
\rightarrow T_{0}e^{-3\omega_{1}}$~~if~$\omega_{0}=\omega_{1} $~~\\

~~{\it(d)}~~$ C_{v}\rightarrow
\frac{S_{0}}{\omega_{0}-\omega_{1}}$~~,~${v_{s}}^{2} \rightarrow
\omega_{0}-\omega_{1} $~~\\

The limiting behaviour shows that for realistic fluid we must have
~$ \omega_{0}>\omega_{1} $~~. Here the energy density has correct
behaviour, pressure is positive throughout the evolution
(approaches to zero) with finite sound velocity. Also the third
law of thermodynamics is satisfied as temperature approaches to
zero with volume goes to infinity (see Fig.1(b)). Further as
~$C_{v}>0 $~so the evolution from large temperature to ~$T=0$~ is
a thermodynamically stable transition without any critical point.
When ~$\omega_{0}<\omega_{1}$~  , then pressure becomes negative
at an intermediate stage of evolution and finally approaches to
zero $-$ a possibility of accelerating phase of the universe. Also
as ~$C_{v}$~ changes sign (shown in Fig.1(d))~so there is a
critical point indicating a phase transition of the
thermodynamical system. For ~$\omega_{0}=\omega_{1}$~ , the
pressure cannot be negative at any stage of the evolution, the
temperature approaches a finite non-zero value but heat capacity
approaches infinite value as volume increases infinitely. Lastly,
it should be noted that from equations (10) and (11) ~$z$~ can be
obtained in principle as a function of ~$\rho$~ and ~$T$~ and then
substituting in (10) it is possible to have ~$\rho$~ (and hence
~$p$~) as a function of ~$T$~ and hence all other thermodynamical
variables can be expressed as
a functions of temperature.\\

\textbf{Case II
:~$\omega(z)=\omega_{0}+\frac{\omega_{1}z}{(1+z)}$~}

Corresponding to this equation of state,~$\rho$~ can be obtained
from the energy conservation equation (9)as

\begin{equation}
\rho(z)=\rho_{0}e^{\frac{3\omega_{1}}{1+Z}}{(1+z)}^{3(1+\omega_{0}+\omega_{1})}
\end{equation}

\begin{figure}
\includegraphics[height=1.5in]{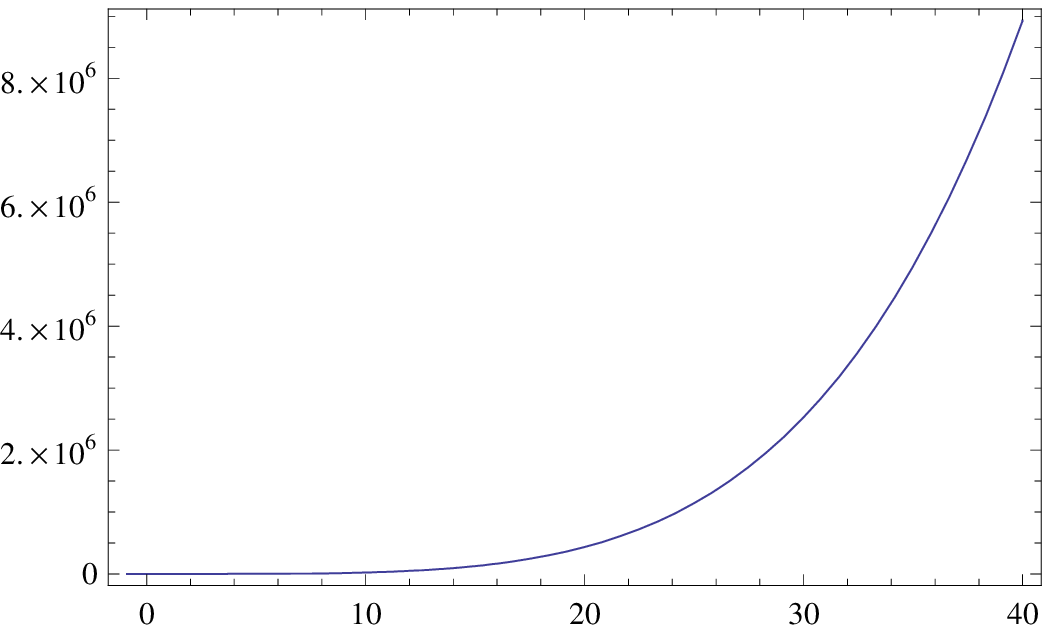}~~~
\includegraphics[height=1.5in]{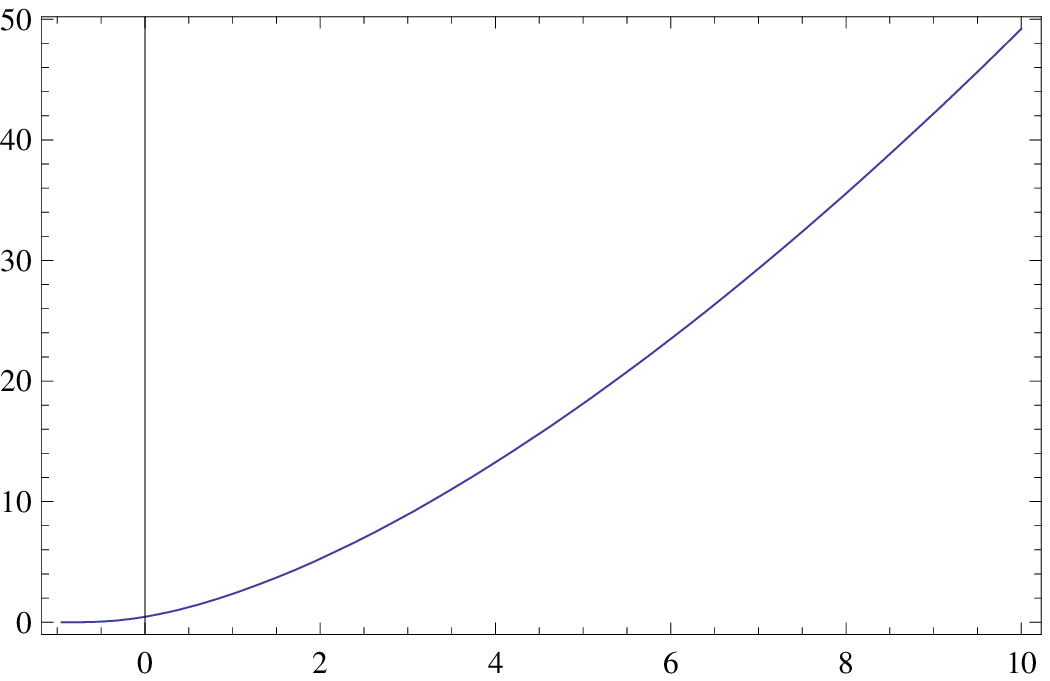}~~~
\\
\vspace{1mm}
Fig.2(a)~~~~~~~~~~~~~~~~~~~~~~~~~~~~~~~~~~~~~~~~~~~~~~~~Fig.2(b)\\

\includegraphics[height=1.5in]{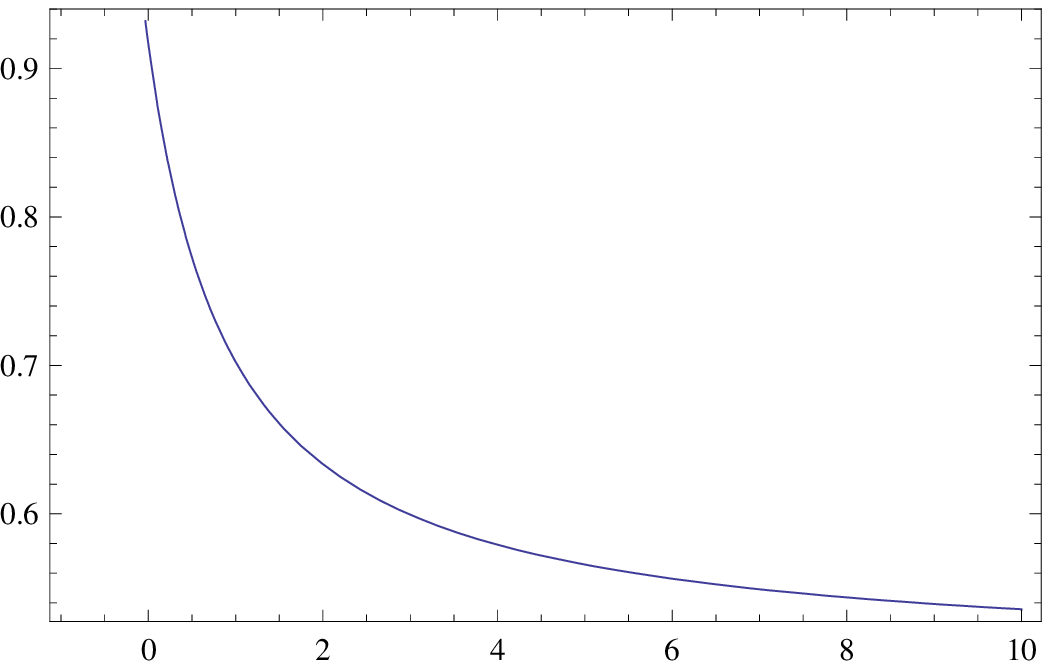}~~~
\includegraphics[height=1.5in]{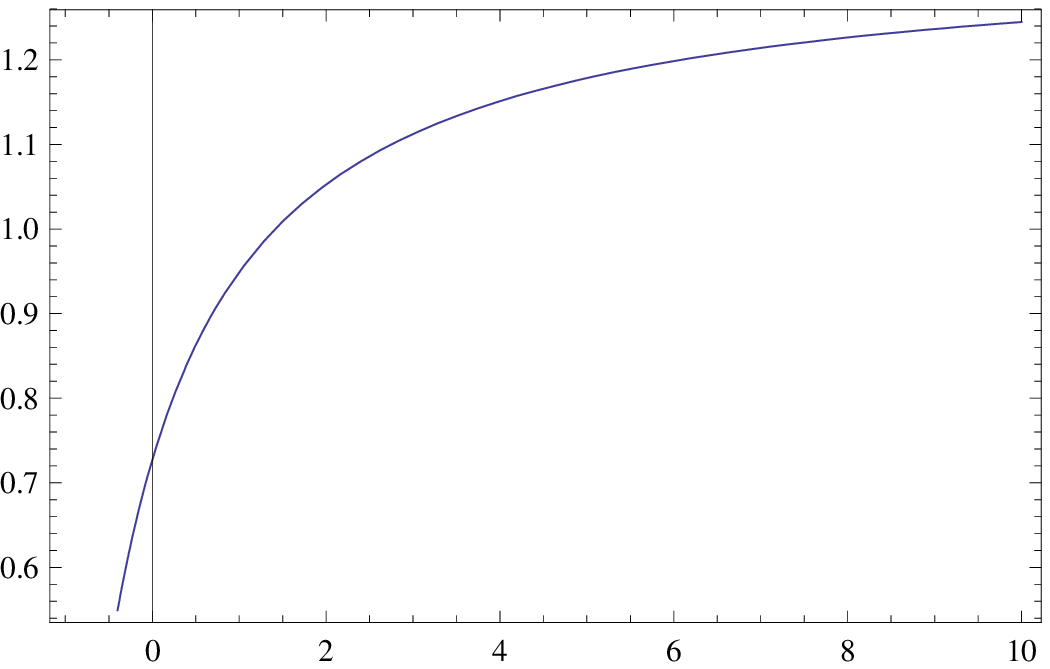}~~~
\\
\vspace{1mm}
Fig.2(c)~~~~~~~~~~~~~~~~~~~~~~~~~~~~~~~~~~~~~~~~~~~~~~~~Fig.2(d)\\

\vspace{5mm} Fig. 2(a),2(b),2(c) and 2(d) shows the variation of
the pressure,temperature ,velocity of sound and heat capacity with
the red shift factor (given by eq.(14,15,16,17))respectively where
$\rho_{0}=1,~S_{0}=1$ and ~$\omega_{1}<0 $~~ , with
~$\omega_{0}=1,\omega_{1}=(-.5)$~ \hspace{1cm} \vspace{6mm}
\end{figure}

Due to the adiabatic nature of the thermodynamical system
temperature can be obtained from equation(4) as a function of z
namely

\begin{equation}
T(z)=T_{0}e^{\frac{3\omega_{1}}{1+Z}}{(1+z)}^{3(\omega_{0}+\omega_{1})-1}(1+\omega_{0}+z(1+\omega_{0}+\omega_{1}))
\end{equation}

The expressions for the squared speed of sound ~$v_{s}^{2}$~ and
the heat capacity ~$ C_{v}$~ are given by

\begin{equation}
{v_{s}}^{2}(z)=\omega+\frac{\omega_{1}}{3(1+z)(1+\omega)}
\end{equation}

and

\begin{equation}
 C_{v}(z) = \frac {3S_{0}}{1+z}\left[\frac{3
\omega}{1+z}+ \frac{\omega_{1}}{(1+\omega)(1+z)^{2}}\right]^{-1}
\end{equation}

\textbf{Limiting Values : }\\

As $ a \rightarrow 0$ i.e. $z \rightarrow \infty $we can find~(
from fig.2) $\rho \rightarrow \infty $,$p\rightarrow \infty$,$T
\rightarrow \infty $,${v_{s}}^{2} \rightarrow
(\omega_{0}+\omega_{1})$ and $C_{v} \rightarrow
\frac{S_{0}}{\omega_{0}+\omega_{1}}$~~.In this limiting case
$\omega \rightarrow \ {\omega_{0}+\omega_{1}} $~~\\

Also if $ a \rightarrow \infty $ i.e.$z \rightarrow -1$ we have ~$
\omega \rightarrow \infty $~if~$\omega_{1}<0$ and ~$\omega
\rightarrow -\infty $~if~$\omega_{1}>0$~~ \\

~~{\it(i)}$\rho \rightarrow 0 $ if $ \omega_{1}<0 $~and ~~$\rho
\rightarrow \infty $ if $ \omega_{1}>0 $~\\

{\it(ii)}$p\rightarrow 0 $ if $ \omega_{1}<0 $~~and
~~$p\rightarrow {\infty} $ if $\omega_{1}>0$\\

{\it(iii)}~$T \rightarrow 0$ if $\omega_{1}<0$,~~$T \rightarrow
\infty $ if $\omega_{1}>0$~~\\

{\it(iv)}$ C_{v}\rightarrow 0$~~,~${v_{s}}^{2} \rightarrow \infty
$~if~$\omega_{1}<0$ and ~${v_{s}}^{2} \rightarrow
-\infty $~if~$\omega_{1}>0$~~\\

$\omega_{1}<0$ but $\omega_{0}>|\omega_{1}|$ then this equation of
state is purely a decelerating model of the universe. But if
$\omega_{0}<|\omega_{1}|$ then the universe changes from
accelerating phase to decelerating phase. The velocity of sound
has a finite value at early epoch and then approaches to infinite
with the evolution of the universe while heat capacity starting
from a finite value decreases to zero. The thermodynamical system
obey the third law as ~$T\rightarrow 0$~ when volume becomes
infinite. However , for ~$\omega_{1}>0$~, though the universe
evolve from decelerating phase to an accelerating phase, yet it is
not a realistic model of the universe because $\rho$ and $T$
decreases in both the limits and square of the velocity of sound
becomes negative at some intermediate instant and approaches
~$-\infty$~ as volume becomes very large. \\

\textbf{Case-III:~$\omega(z)
=\omega_{0}+\frac{\omega_{1}z}{(1+z)^{2}}$}

Corresponding to this adiabatic equation of state the relevant
thermodynamical quantities are given by

\begin{equation}
\rho(z)=\rho_{0}(1+z)^{3(1+\omega_{0})}
e^{-\frac{3\omega_{1}(2z+1)}{2(1+z)}^{2}}
\end{equation}

\begin{equation}
T(z)=T_{0}(1+\omega)(1+z)^{3
\omega_{0}}e^{-\frac{3\omega_{1}(2z+1)}{2(1+Z)}^{2}}
\end{equation}

\begin{equation}
{v_{s}}^{2}(z)=\omega_{0}+\frac{\omega_{1}}{(1+z)^{2}}\left[~z+\frac{(1-z)}{3(1+\omega)}\right]
\end{equation}

\begin{equation}
 C_{v}(z) = \frac {3S_{0}}{1+z}\left[\frac{3 \omega}{1+z}+
\frac{\omega_{1}(1-z)}{(1+\omega)(1+z)^{3}}\right]^{-1}
\end{equation}

\begin{figure}
\includegraphics[height=1.5in]{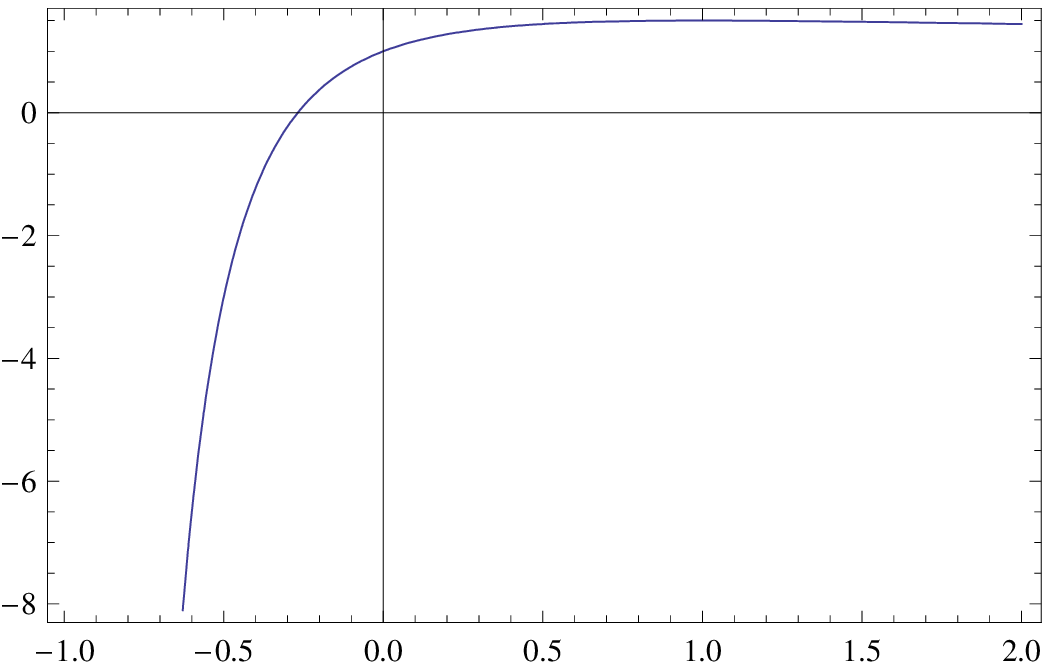}~~~
\includegraphics[height=1.5in]{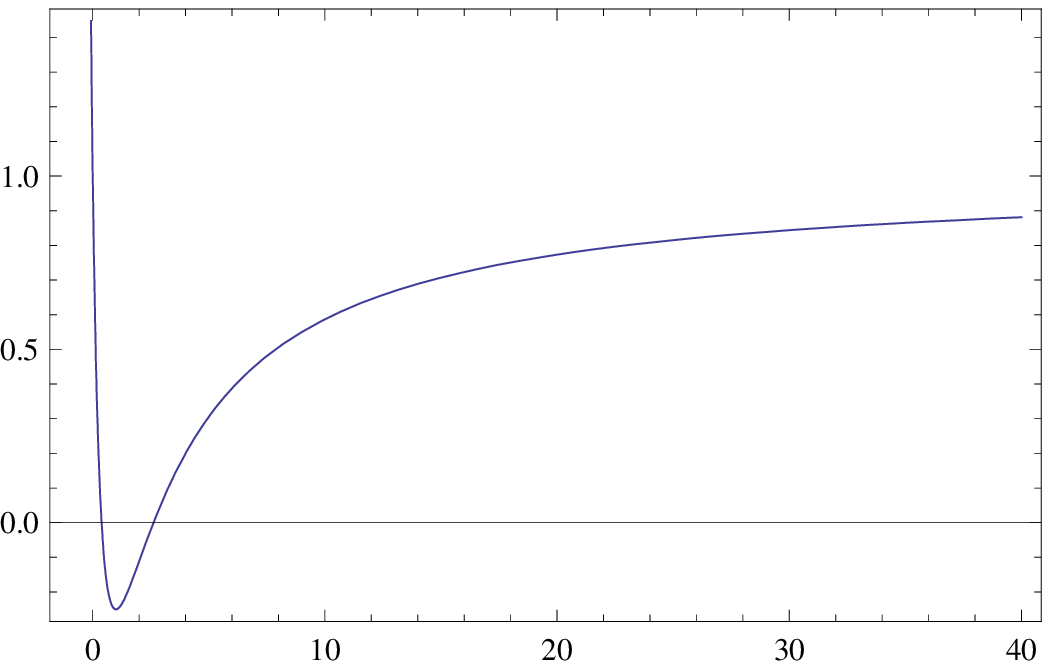}~~~
\\
\vspace{1mm}
Fig.3(a)~~~~~~~~~~~~~~~~~~~~~~~~~~~~~~~~~~~~~~~~~~~~~~~~Fig.3(b)\\

\vspace{5mm} Fig. 3(a),3(b) shows the variation of the equation of
state with the red shift factor when  ~$\omega_{1}>0 $~~and
~$\omega_{1}<0 $~~ respectively. For
Fig.3(a)~$\omega_{1}=2,\omega_{0}=1$~ and for Fig.
3(b)~$\omega_{1}=(-5),\omega_{0}=1$~\hspace{1cm} \vspace{6mm}
\end{figure}

\begin{figure}
\includegraphics[height=1.5in]{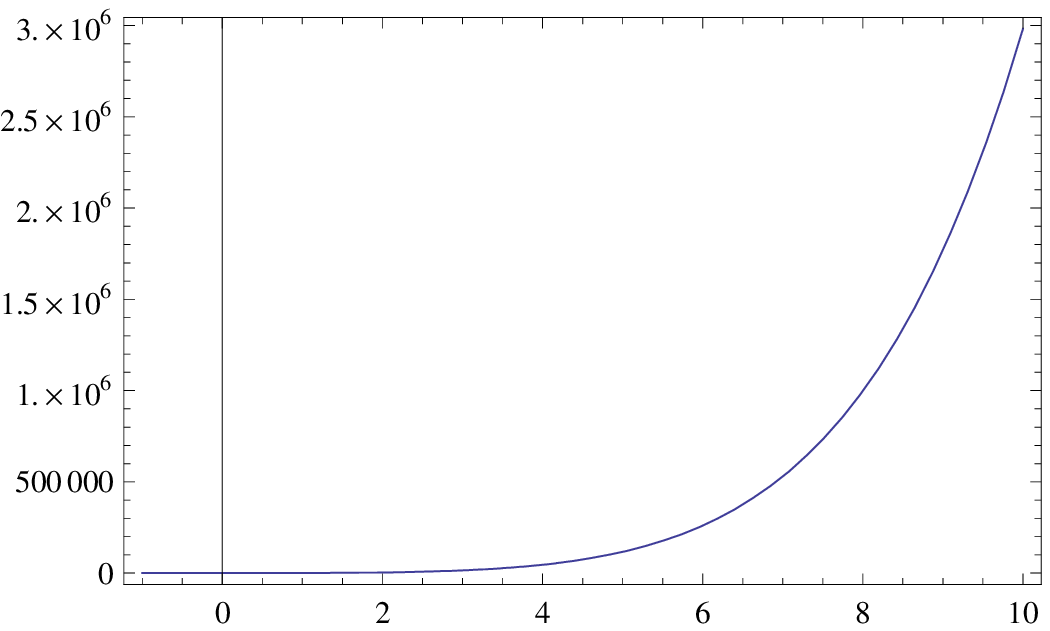}~~~
\includegraphics[height=1.5in]{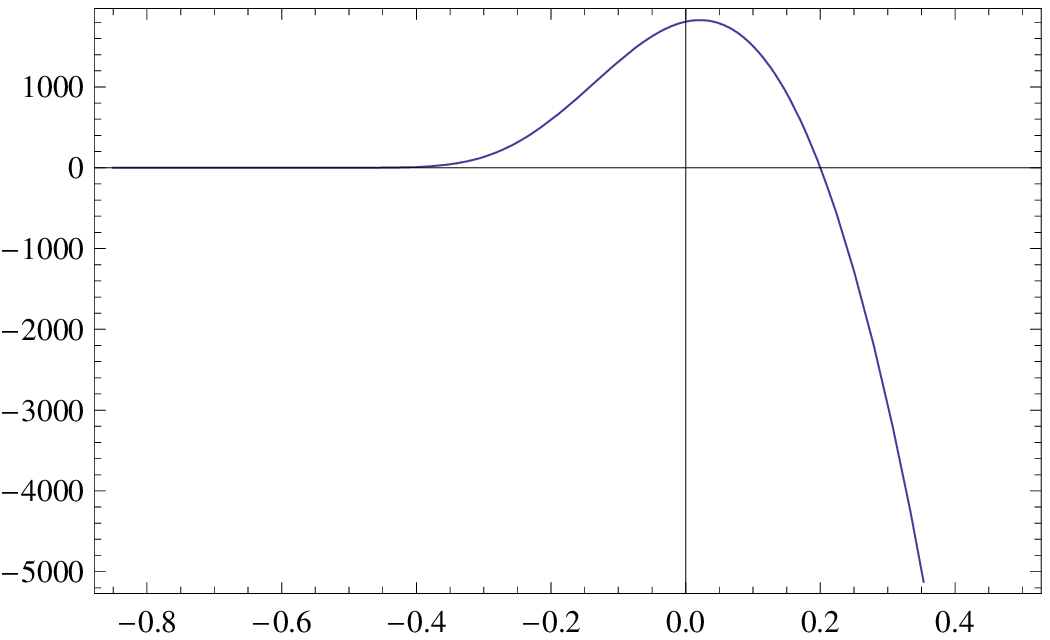}~~~
\\
\vspace{2mm}
Fig.4(a)~~~~~~~~~~~~~~~~~~~~~~~~~~~~~~~~~~~~~~~~~~~~~~~~Fig.4(b)\\
\includegraphics[height=1.5in]{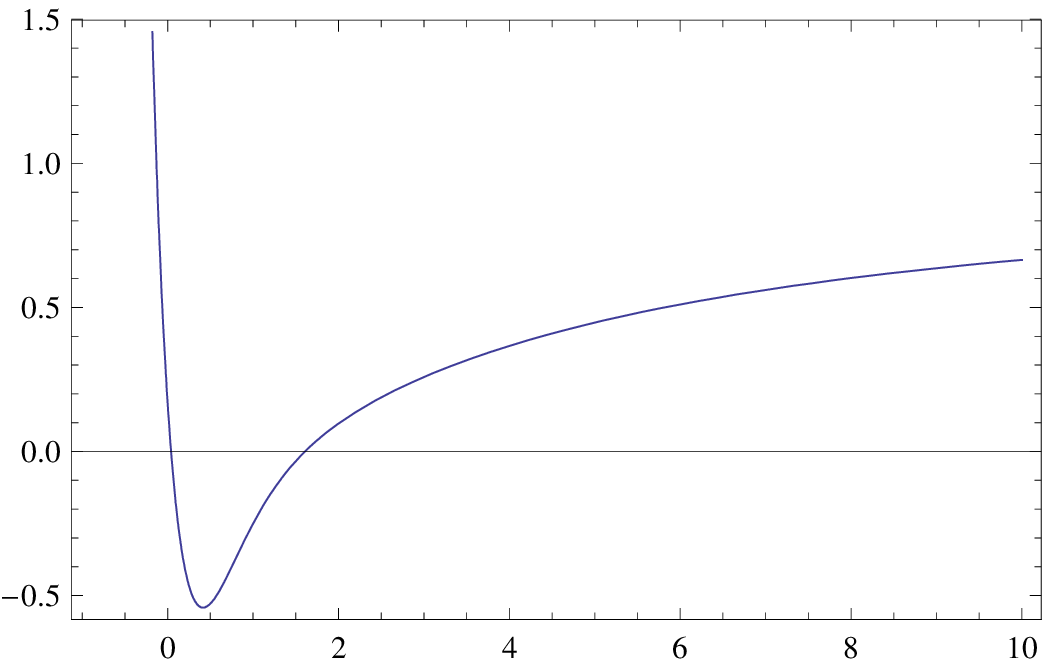}~~~
\includegraphics[height=1.5in]{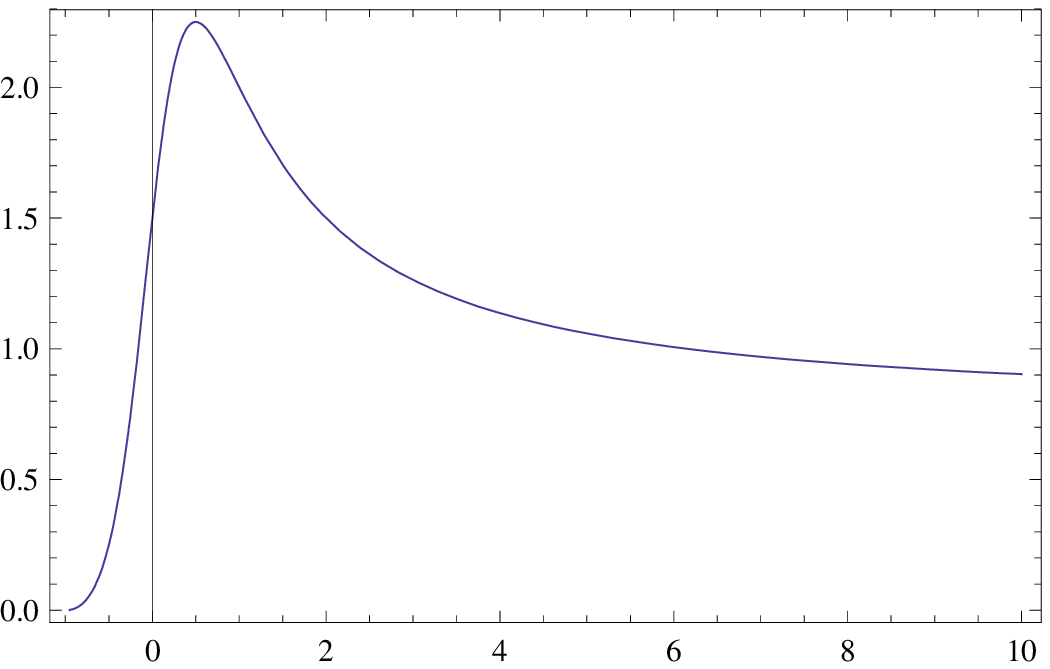}~~~
\\
\vspace{.75mm}
Fig.4(c)~~~~~~~~~~~~~~~~~~~~~~~~~~~~~~~~~~~~~~~~~~~~~~~~Fig.4(d)\\

\vspace{5mm} Fig. 4(a),4(b),4(c) and 4(d) shows the variation of
the energy density, pressure, velocity of sound and heat capacity
with the red shift factor (given by eq. (18, 19, 20, 21))
respectively, where $\rho_{0}=1,~S_{0}=1$ and ~$\omega_{1}<0 ,
\omega_{0}>0$~~with~$\omega_{0}=1, \omega_{1}=(-5)$~ \hspace{1cm}
\vspace{6mm}
\end{figure}

\textbf{Limiting Values : }\\

As $ a \rightarrow 0$ i.e. $z \rightarrow \infty $we can find~(
from fig.3) $\rho \rightarrow \infty $,~$p\rightarrow \infty$,~$T
\rightarrow \infty $,~${v_{s}}^{2} \rightarrow \omega_{0}$ and
$C_{v} \rightarrow \frac{S_{0}}{\omega_{0}}$~~.In this limiting
case $\omega \rightarrow \ {\omega_{0}} $~~\\

Also if $ a \rightarrow \infty $ i.e.$z \rightarrow -1$ we have ~$
\omega \rightarrow \infty $~if~$\omega_{1}<0$ and ~$\omega
\rightarrow -\infty $~if~$\omega_{1}>0$~~ \\

~~{\it(i)}~~$\rho \rightarrow 0 $ if $ \omega_{1}>0 $~and ~~$\rho
\rightarrow \infty $ if $ \omega_{1}<0 $~\\

{\it(ii)}$p\rightarrow \infty $ if $ \omega_{1}<0 $~~and
~~$p\rightarrow {-\infty} $ if $\omega_{1}>0$\\

{\it(iii)}~$T \rightarrow 0$ if $\omega_{1}>0$,~~$T \rightarrow
\infty $ if $\omega_{1}<0$~~\\

{\it(iv)}$ C_{v}\rightarrow 0$~~,~${v_{s}}^{2} \rightarrow \infty
$~if~$\omega_{1}<0$ and ~${v_{s}}^{2} \rightarrow
-\infty $~if~$\omega_{1}>0$~~\\

for this equation of state if ~$\omega_{1}>0$~~then initially the
universe has a decelerating phase (if ~$\omega_{0}>0$~~) and then
it has accelerated expansion[in Fig.3a] while for negative
~$\omega_{1}$~( with ~$\omega_{0}>0$~)~ the universe changes from
decelerating phase to an accelerating phase then again there is a
deceleration provided ~$|\omega_{1}|>4|\omega_{0}|$~~[in Fig.3b].
Although the third law of thermodynamics is satisfied for
~$\omega_{1}<0$~, the thermodynamical system is not a realistic
one as velocity of sound become imaginary at an instant. On the
other hand for ~$\omega_{1}>0$~~ the density, pressure and
temperature starting from an infinite value reach a minimum value
and then again increases unboundedly. The heat capacity starting
from a finite value gradually approaches to zero with the
evolution of the universe . Note that ,although we have a
deceleration$-$acceleration$-$deceleration phase for
~$\omega_{1}<0$~~ and ~$\omega_{0}>0$~~ but the only drawback is
the imaginary sound speed at some finite red
shift.\\

\textbf{case-IV :~$$\omega(z)=\omega_{0}+\omega_{1}z,~~if~z<1$$
~$$\omega_{0}+\omega_{1}~~if~z\geq1$$}

This equation of state represent a decelerating model of the
universe throughout the evolution if ~$\omega_{0}>\omega_{1}$~
while for ~$\omega_{0}<\omega_{1}$~ there will be a transition
from deceleration to acceleration at
~$z=-\frac{\omega_{0}}{\omega_{1}}$~. The physical parameters
characterizing the thermodynamical system are given by

\begin{equation}
\rho(z)=\rho_{0}e^{3\omega_{1}z}{(1+z)}^{3(1+\omega_{0}-\omega_{1})}
~~ifz<1~~~~~~~and~~~~~~\rho_{0}{(1+z)}^{3(\omega_{0}+\omega_{1})}~~if~z\geq1
\end{equation}

\begin{equation}
T(z)=T_{0}e^{3\omega_{1}z}{(1+z)}^{(\omega_{0}-\omega_{1})}(1+\omega_{0}+\omega_{1}z)
~~ifz<1~~~~~~~and~~~~~~T_{0}{(1+z)}^{3(\omega_{0}+\omega_{1})}~~if~z\geq1
\end{equation}

\begin{equation}
{v_{s}}^{2}(z)=\omega+\frac{\omega_{1}(1+z)}{3(1+\omega)}
~~ifz<1~~~~~~~and~~~~~~\omega_{0}+\omega_{1}~~if~z\geq1
\end{equation}

\begin{equation}
C_{v}(z)=\frac {3S_{0}}{1+z}{\left[~{\frac{3 \omega}{1+z}+
\frac{\omega_{1}}{1+\omega}}~\right]}^{-1}~~ifz<1~~~~~~~and~~~~~~\frac
{S_{0}}{{1+z}^{3}}~~if~z\geq1
\end{equation}

\textbf{Limiting Values : }\\

For the volume of the universe approaches zero or infinity the
limiting value of these thermodynamical  parameters are given by
as follows:\\

~~{\bf(I).}~~$ a \rightarrow 0$ i.e. $z \rightarrow \infty $:\\

{\it(i)}$\rho \rightarrow \infty $,~$p\rightarrow \infty$,~$T
\rightarrow \infty$ ~if ~$(\omega_{0}+\omega_{1})>0$~\\

{\it(ii)}$\rho \rightarrow 0 $,~$p\rightarrow 0$,~$T \rightarrow
0$~if ~$(\omega_{0}+\omega_{1})< 0$~\\

{\it(iii)}$\rho \rightarrow \rho_{0} $,~$p\rightarrow 0$,~ $T
\rightarrow p_{0}$ ~if ~$(\omega_{0}+\omega_{1})= 0$~\\

{\it(iv)}${v_{s}}^{2} \rightarrow (\omega_{0}+\omega_{1})$ ,
$C_{v} \rightarrow 0$~~ and $\omega \rightarrow \ {\omega_{0}+\omega_{1}} $~~\\

~~{\bf(II).}~$ a \rightarrow \infty $ i.e.$z \rightarrow -1$ :\\

{\it(i)}~~$\rho ,~p ,~T\rightarrow 0 $ if $
\omega_{1}<\omega_{0} $~~\\

{\it(ii)}$\rho,~p,~T \rightarrow \infty $~if~ $
\omega_{0}<\omega_{1}<1+\omega_{0} $~~\\

{\it(iii)}${v_{s}}^{2} \rightarrow (\omega_{0}-\omega_{1})$ ~, ~$
C_{v}\rightarrow \frac{S_{0}}{\omega_{0}-\omega_{1}}$~and~~$
\omega \rightarrow {\omega_{0}-\omega_{1}} $~\\

It is clear that ~$\omega_{0}<\omega_{1}$~ is not physically
reasonable as velocity of sound become imaginary for infinitely
large volume. For ~$\omega_{0}>\omega_{1}$~ the third law of
thermodynamics is satisfied, velocity of sound has finite non
zero value and heat capacity gradually increases to a finite value
throughout the evolution.

\section{\normalsize\bf{Validity of laws of thermodynamics}}

Suppose the line element (see equation(6)) for FRW model is
written in the form
\begin{equation}
ds^{2}=h_{ab}dx^{a}dx^{b}~+~{\tilde{r}}^{2}d{{\Omega}_{2}}^{2}
\end{equation}
where $\tilde{r}=ar$ is the area radius (or geometrical radius)
and ~$h_{ab}=~diag(-1,\frac{a^{2}}{(1-kr^{2})})$~is the metric on
the 2~dimensional hyper surface ($x^{0}=t,x^{1}=r$ ) .Then the
dynamical apparent horizon which is a marginally trapped surface
with vanishing expansion, is given by the relation

\begin{equation}
h^{ab}d_{a}\tilde{r}d_{b}\tilde{r}=0
\end{equation}

i.e. $$1-{R_{A}}^{2}\left(H^{2}+\frac{k}a^{2}\right)=0$$

The above relation shows that for flat space($k=0$)~$R_{A}$,the
radius of the apparent horizon coincide with the Hubble horizon.
           The cosmological event horizon is defined as

\begin{equation}
R_{E}=a\int^{\infty}_{t}\frac{dt}{a}=a\int^{\infty}_{a}\frac{da}{Ha^{2}}
\end{equation}

 The thermodynamical quantities namely entropy and temperature have the following
expressions at the horizon

\begin{equation}
S_{I}=\frac{A}{4G}=\frac{\pi{R_{I}}^{2}}G~~,~~T_{I}=\frac{1}{2\pi
R_{I}}
\end{equation}

where $I = A$ or $E$ according as we consider apparent or
cosmological event horizon.\\

 The  first law of thermodynamics on a horizon
(apparent or cosmological event horizon) can be written as

\begin{equation}
-dE_{I}=T_{I}dS_{I}
\end{equation}

 Where~$-dE_{I}$~ is the amount of energy crossing the
horizon during the time interval ~$dt$ .For the present ~$4D$ FRW
space time the expression for ~$ dE_{I}$~ is given by

\begin{equation}
-dE_{I}=4\pi{R_{I}}^{3}H(\rho+p)dt~,
\end{equation}

 where $\rho$ and $p$ are the energy density
and thermodynamic pressure of  the matter in the universe bounded
by the horizon and the equation of state is given by (as in the
previous section ) ~$p$=$\omega\rho$ where ~$\omega$~ is a
function of the red shift variable ~$z$. Now for the horizon
$R_{I}$ the change of entropy is given by (choosing G=1)

\begin{equation}
dS_{I}=2\pi R_{I}dR_{I}
\end{equation}

 For the apparent horizon $ R_{A} $ using the definition and the
Friedmann equation (7) we have from above (on simplification)

\begin{equation}
d{S_{A}}^2=8{\pi}^{2}{R_{A}}^{4}H(\rho+p)dt
\end{equation}

and hence $$T_{A}dS_{A}=4\pi{R_{A}}^{3}H(\rho+p)dt$$

So from equation (31) we have
$$-dE_{A}=T_{A}dS_{A}$$

Thus the first law of thermodynamics is obeyed at the apparent
horizon.The validity of this law does not depend on the equation
of state of the fluid in the universe bounded by the apparent
horizon. \\
On the other hand, the radius of the event horizon cannot be
obtained in a close form (only in the integral form given by
equation ~(28) for arbitrary equation of state and as a result the
right hand side of equation (7) can no longer be simplified
further . Hence no conclusion can be made about the validity of
the first law of thermodynamics  at the cosmological event
horizon. \\
To examine the validity of the second law at the horizon (apparent
or event) we first study the entropy enveloped by the
horizon.Using the Gibb's equation [11] one can relate the entropy
of the universe inside the horizon to its energy and pressure in
the horizon as

\begin{equation}
TdS_{m}=dE{m}+pdV
\end{equation}

Where $V=\frac{4\pi {R_{I}}^{3}}{3}$  is the volume bounded by the
horizon, $E_{m}$ =$\frac{4\pi {R_{I}}^{3}\rho}{3}$  is the energy
inside the horizon and ~ $S_{m}$ ~ is the entropy of the matter
distribution inside the horizon. We now consider the apparent
horizon as the boundary of the universe.For thermodynamical
equilibrium we can choose the temperature ~ $T_{A}$ ~ on the
horizon as the temperatue of the matter inside ~$R_{A}$~. Then
from the above Gibb's law(i.e. equation (34))[12]

\begin{equation}
\frac{dS_{m}}{dt} = 2\pi R_{A}q(1+q)
\end{equation}

In deriving this  equation we have used the Friedmann equation (8)
and energy conservation relation(9).Here
 ~ $q=-1-\frac{\dot{H}}{H^{2}}$ ~ is the usual deceleration
parameter.In the similar way from equation (33) we have

\begin{equation}
\frac{dS_{A}}{dt} = 2\pi R_{A}(1+q)
\end{equation}

and hence

\begin{equation}
\frac{d(S_{m}+S_{A})} {dt} = 2\pi R_{A}(1+q)^{2}\geq 0
\end{equation}

Thus the total entropy of the matter distribution inside the
(apparent) horizon and the entropy of the horizon always increases
with the evolution of the universe i.e. the generalized second law
of thermodynamics is valid for the Universe bounded by the
apparent horizon. However,  the entropy of the matter distribution
inside the apparent horizon increases for decelerating phase of
the universe while the entropy decreases in the accelerating
phase.\\
As we have mentioned earlier that there is no explicit analytic
form of $R_{E}$ so it is not possible to obtain an analytic
expression for both ~$\frac{dS_{m}}{dt}$~ and
~$\frac{dS_{E}}{dt}$~ and hence we cannot draw any conclusion
about increase or decrease of the entropy with the evolution of
the universe.So no positive conclusion is possible for the
validity of the thermodynamical
laws at the cosmological event horizon.\\

\section{\normalsize\bf{Discussions}}

In the present paper we consider the general thermodynamics of the
universe filled with perfect fluid having equation of state
~$p=\omega\rho$. Here $\omega$ is chosen as a function of the red
shift variable  {\it z }. Four different choices for $\omega$ are
taken and thermodynamics with asymptotic limits have been
discussed.These choices of $\omega$ are recently shown to be in
good agreement with current observations in different ranges of
{\it z }.The third choice of $\omega$  namely that of
Jassal-Bagala-Padmanavan seems to be interesting because it shows
a transition from deceleration to acceleration and then again
deceleration as expected from observation. \\
          Next section deals with validity of
the thermodynamical laws for the universe filled with perfect
fluid (~$p=\omega\rho,~\omega$ is variable) and is bounded by
apparent or cosmological event horizon. As the event horizon
radius can not be evaluated in closed form (only in an integral
form) so it is not possible to infer about the validity of the
thermodynamical laws. However for the apparent horizon,both first
and second law of thermodynamics are obeyed for general $\omega$
(without specifying it). For the validity of the second law we
have started with the Gibb's equation and calculated the rate of
change of the sum total of the entropy of the matter and the
entropy of the boundary which is always positive. This is known as
generalized second law of thermodynamics.For future work it will
be interesting to consider holographic model of dark energy so
that an analytic expression for the radius of the event horizon
can be obtained and then it will be possible to examine the
validity of thermodynamical laws at the event horizon.\\

{\bf Acknowledgement:}\\

A part of the work has been carried out during a visit to
IUCAA.The authors are thankful to IUCAA, Pune ,India for warm
hospitality and facility of researches.Also the authors are
thankful to Dr. U.Debnath (Bengal Engineering and Science
University ) for helping in preparing the manuscript.\\

{\bf References:}\\
\\
$[1]$ S.W.Hawking, \textit{Commun.Math.Phys} \textbf{43} 199 (1975).\\
$[2]$ J.D.Bekenstein, \it{Phys. Rev. D} {\bf 7} 2333 (1973).\\
$[3]$ J.M.Bardeen , B.Carter and S.W.Hawking, {\it
Commun.Math.Phys }
{\bf 31} 161 (1973).\\
$[4]$ T.Jacobson, \it {Phys. Rev Lett.} {\bf 75} 1260 (1995) \\
$[5]$ T.Padmanabhan, \it {Class.
Quantum Grav} {\bf 19} 5387 (2002); \it{Phys.Rept} {\bf 406} 49 (2005)\\
$[6]$ R. G. Cai and S. P. Kim, {\it JHEP} {\bf 02} 050 (2005).\\
$[7]$ M. Akbar and R.G. Cai ,\it{ Phys. Lett. B } {\bf 635} 7
(2006) ; A. Paranjape,S. Sarkar and T. Padmanavan, {\it Phys. Rev.
D} {\bf 74} 104015 (2006).\\
$[8]$ V.B. Johri and P.K. Rath, {\it International J. of Modern Phys } {\bf 16 } 1581 {2007}.\\
$[9]$  Y.S. Myung arXiv:0812.0618[gr-qc]\\
$[10]$ Y. Gong,B. Wang and A. Wang, {\it JCAP } {\bf 01} 024 (2007).\\
$[11]$ G. Izquierdo and D. Pavon, {\it Phys. Lett. B} {\bf 633} 420 (2006).\\
$[12]$ B. Wang,Y. Gong,E. Abdalla, \it{Phys. Rev. D} {\bf 74} 083520 (2006).\\

\end{document}